\documentclass[a4paper]{jpconf}
\usepackage{graphicx}
\usepackage{float}
\begin{document}
\title{Measurement of Di-electron Continuum in Au+Au collisions  at $\sqrt{s_{NN}}=200$~GeV by PHENIX using the Hadron Blind Detector}

\author{Sky D. Rolnick for the PHENIX Collaboration}

\address{UC Riverside}

\ead{skymeson@gmail.com}

\begin{abstract}
Dielectrons play an important role in heavy ion collisions since they are produced during all stages of the collision and interact only electromagnetically. The PHENIX experiment was upgraded with the Hadron Blind Detector which provides excellent background rejection by removing pairs from both partially reconstructed Dalitz decays and photon conversions. In these proceedings we will report on the results obtained from 2009 data in $p$+$p$ collisions using the HBD as well as the Au+Au results obtained in 2010. 

\end{abstract}

\section{Introduction}

Dileptons are very useful probes for studying many interesting puzzles in Heavy Ion Physics. They are produced during all stages of the collision and do not interact strongly with the medium.  The dielectron continuum offers a rich and diverse region of physics to explore. It allows us to study such effects as chiral symmetry restoration or possible medium modifications of vector mesons. There are several known sources which contribute to the dielectron continuum including thermal radiation, direct photons, vector meson decays, and open charm and bottom yields.  The cleanness of the dielectron measurements, compared to hadronic measurements, provides probes into multi-component flow and energy loss mechanisms and will provide checks on a variety of production mechanisms. 

Previous measurements made by PHENIX in Au+Au collisions at $\sqrt{s_{NN}}=200$~GeV have shown strong enhancement of electron pairs compared to the expected hadronic cocktail in the low mass region $m_{ee}=0.15-0.75$~GeV/$c^2 $  by a factor of $4.7\pm0.4(stat.)\pm1.5(syst.)$ for central collisions. Such a large enhancement was not expected and has posed a mystery to many theorists.\cite{Gale} The very low signal to background ratio of $\sim1/200$ generates large systematics in the region of interest. To improve on this, a new measurement has been made by PHENIX utilizing the Hadron Blind Detector, which reduces the backgrounds and systematic uncertainty. In this paper, results for $p$+$p$ collisions will be presented along with Au+Au data for three centrality bins and compared to their cocktails.


\section{Hadron Blind Detector}

The Hadron Blind Detector is a novel type of  Cherenkov detector utilizing a triple GEM configuration with a CSI photo-cathode layer operated in pure CF4. The HBD has achieved the highest figure of merit of any gas Cherenkov detector to date $No=322cm^{-1}$ and designed with a very low material budget of just 2.4\% total radiation lengths.\cite{Anderson} It was successfully operated in 2009 $p$+$p$ data and 2010 Au+Au data sets.

\begin{figure}[H]
  \vspace{-3mm}
  \begin{center}
    \includegraphics[keepaspectratio=true,width=0.8\textwidth]{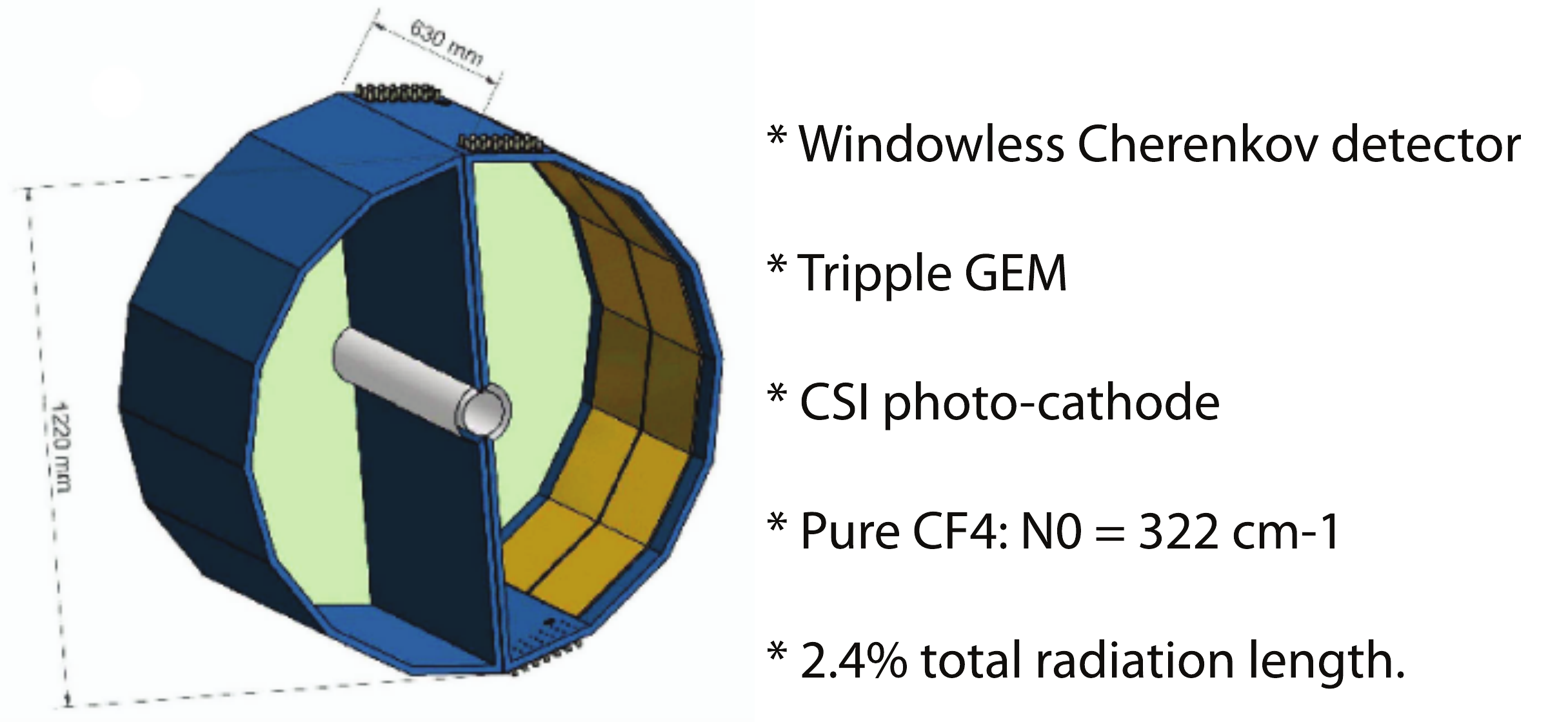}
\end{center}
  \vspace{-3mm}
  \caption{The HBD concept design.}
  \label{fig:fig_hbd_concept}
\end{figure}

The HBD was designed to work by preserving the opening angle of charged particles using the field free region of the PHENIX central arms. Electrons are tagged by the amount of Cherenkov radiation they produce and since heavier meson decays have large opening angles compared to Dalitz decays and conversions,  it is possible to separate these different sources.  Cherenkov light is continuously radiated along the path of the electron such that a small unfocused blob of Cherenkov light is formed on the surface of the detector. The typical photoelectron response is measured to be $\sim$20 p.e. for a single electron and $\sim$40 p.e  for a double electron. In principle the difference in response should distinguish between single and double electrons.\cite{Woody}

Several reconstruction algorithms have been developed and studied with a single electron tracking efficiency of $\sim$90\%.\cite{Watanabe} The main source of background arises from partially reconstructed $\pi^{0}$ Dalitz decays or gamma conversion in which one electron is lost either due to acceptance or tracking inefficiencies. The HBD allows for the tagging and removal of these uncorrelated electron pairs thereby reducing the background.


\section{Hadronic Cocktail}

The hadronic cocktail is generated using EXODUS and filtered through the ideal PHENIX acceptance. This cocktail consists of dielectron decays from known hadronic sources including Dalitz decays of $\pi^{0}$, $\eta$, and $\eta'$ as well as direct decays from vector mesons $\rho, \omega, \phi$ and $J/\psi$. The method for obtaining the cocktail has been outlined in \cite{Adare} and uses a combined fit from neutral and charged pion spectra to a modified Hagedorn function. The $p_{T}$ spectra of the other mesons is derived using $m_{T}$ scaling of the $\pi^{0} $ spectra and normalized using the meson-to-pion ratio at high $p_{T}$. The momentum resolution of PHENIX tracking is taken into account using a smearing technique implemented in EXODUS. The additional material of the HBD was taken into account yielding higher rates of conversion electrons which were modeled using a full Geant simulation of the PHENIX detector. The line shape of the $J/\psi$ was also modified to account for Bremstrahlung radiation produced by the additional material. The difference in magnetic field configurations are reflected in the acceptance of the electron pairs which is also accounted for. The heavy flavor contributions from charm and bottom are estimated using MC@NLO \cite{Frixione} as opposed to PYTHIA \cite{Sjostr} since this found to reproduce dielectron yields, especially at high $p_{T}$. The cocktail used matches very well to $p$+$p$ data over the whole mass range.






\section{Results for $p$+$p$ collisions}


\begin{figure}[ht]
\begin{minipage}{0.5\linewidth}
\includegraphics[width=0.9\textwidth]{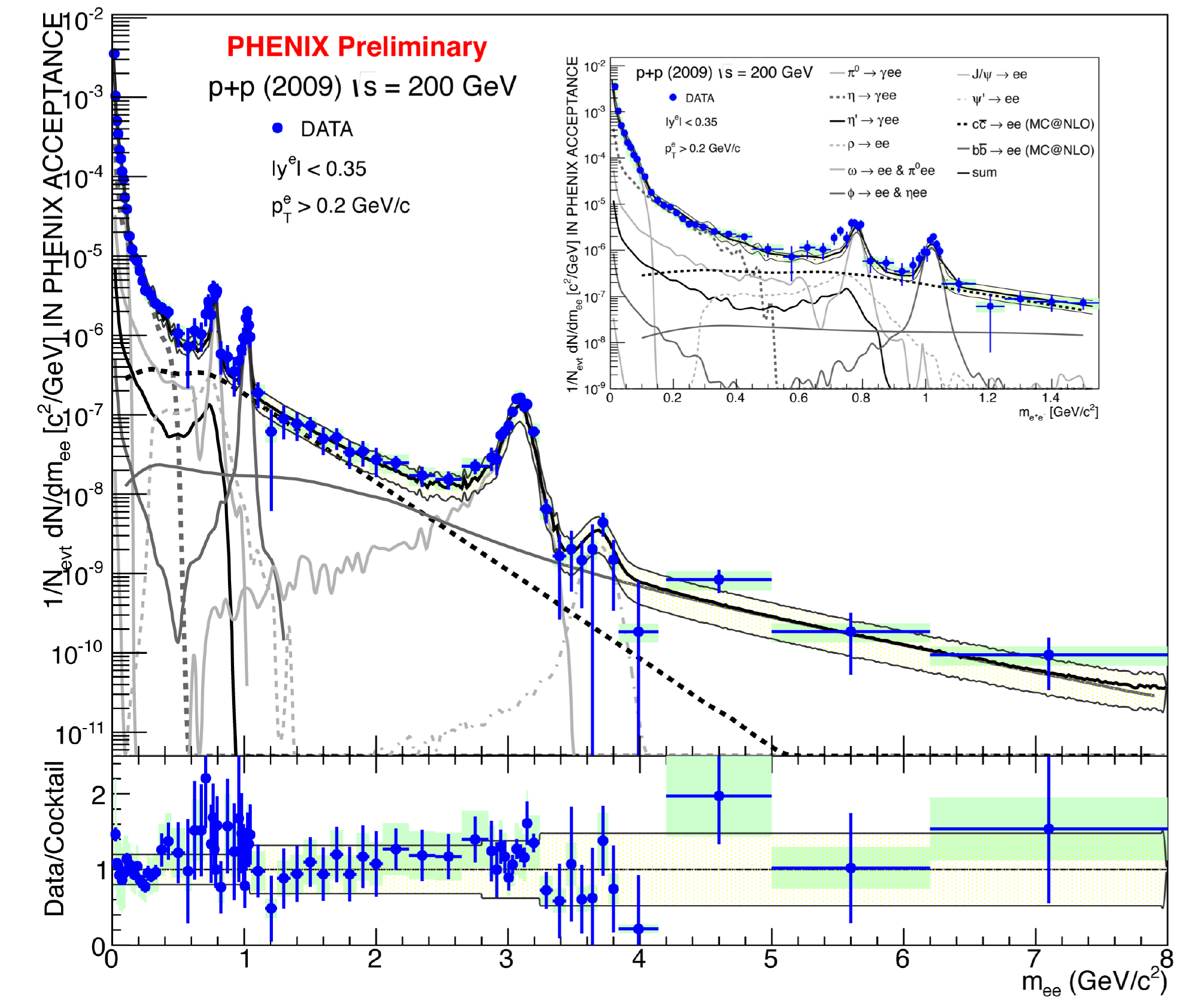}
\end{minipage}
\begin{minipage}{0.5\linewidth}
\includegraphics[width=0.9\textwidth]{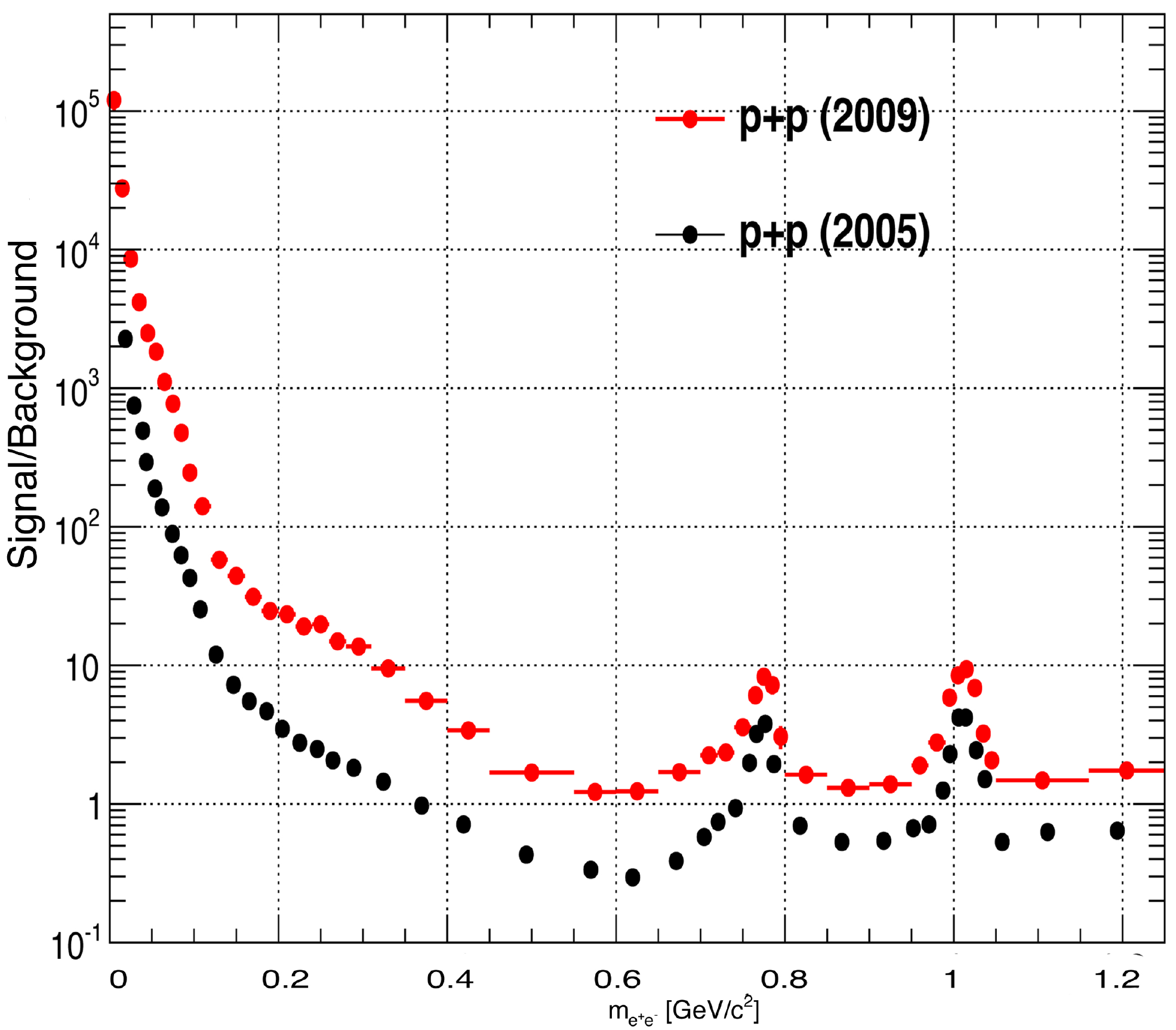}
\end{minipage} 
\caption{\label{fig:compare_pp}The left panel shows the PHENIX dielectron mass spectrum for Run9 $p$+$p$ overlayed with the cocktail shown by a solid line. These results show excellent agreement between data and cocktail (lower panel). The signal to background comparison for 2005 $p$+$p$ data (without HBD) to 2009 $p$+$p$ data (with HBD) is shown in the right panel.}
\end{figure}

The measurements in $p$+$p$ collisions establish the baseline spectra without any medium effects and can be used in comparison to Au+Au results. They also provide a testing ground for understanding the HBD. The S/B ratio is improved by a factor of 5 in the low mass region by using the HBD which demonstrates the effectiveness of the HBD in background rejection. 

Combinatorial background is subtracted using a like-sign subtraction technique which uses electron pairs with the same sign to estimate the correlated and uncorrelated background contributions. Examples of correlated like-sign pairs include cross pairs i.e. $\pi^{0}\rightarrow \gamma \gamma\rightarrow e^{+}e^{-}e^{+}e^{-} $and jet pairs where both electrons are correlated through the jet. Differences in acceptance between like-sign and unlike-sign are corrected with a relative acceptance correction. The main advantage in using like-sign subtraction technique is that combinatorial pairs are produced at the same rate as unlike-sign foreground and the distributions will be absolutely normalized. One drawback of this technique is that the statistical precision is limited compared to mixed event techniques. 



\section{Results for Au+Au collisions}

In Au+Au collisions, the large occupancy rate in the HBD detector presents a challenge especially in central events. The Cherenkov signal begins to get  overwhelmed by other sources of radiation primarily from scintillation but also heavy ionized particles or low momentum electrons looping back into the HBD. In central events, the mean scintillation per pad in the HBD approaches $\sim$7~p.e. and when distributed over the average cluster size of 3 pads per cluster, the response from scintillation becomes comparable to the single electron response. Two techniques have been developed to reduce the background scintillation light including a mean pad subtraction based on centrality and a local background subtraction by looking at nearby pads around the clusters. Using the mean pad subtraction technique, occupancy rates can be reduced to $\sim$15\% for central collisions. 

\begin{figure}[ht]
\begin{minipage}{0.33\linewidth}
\includegraphics[width=0.9\textwidth]{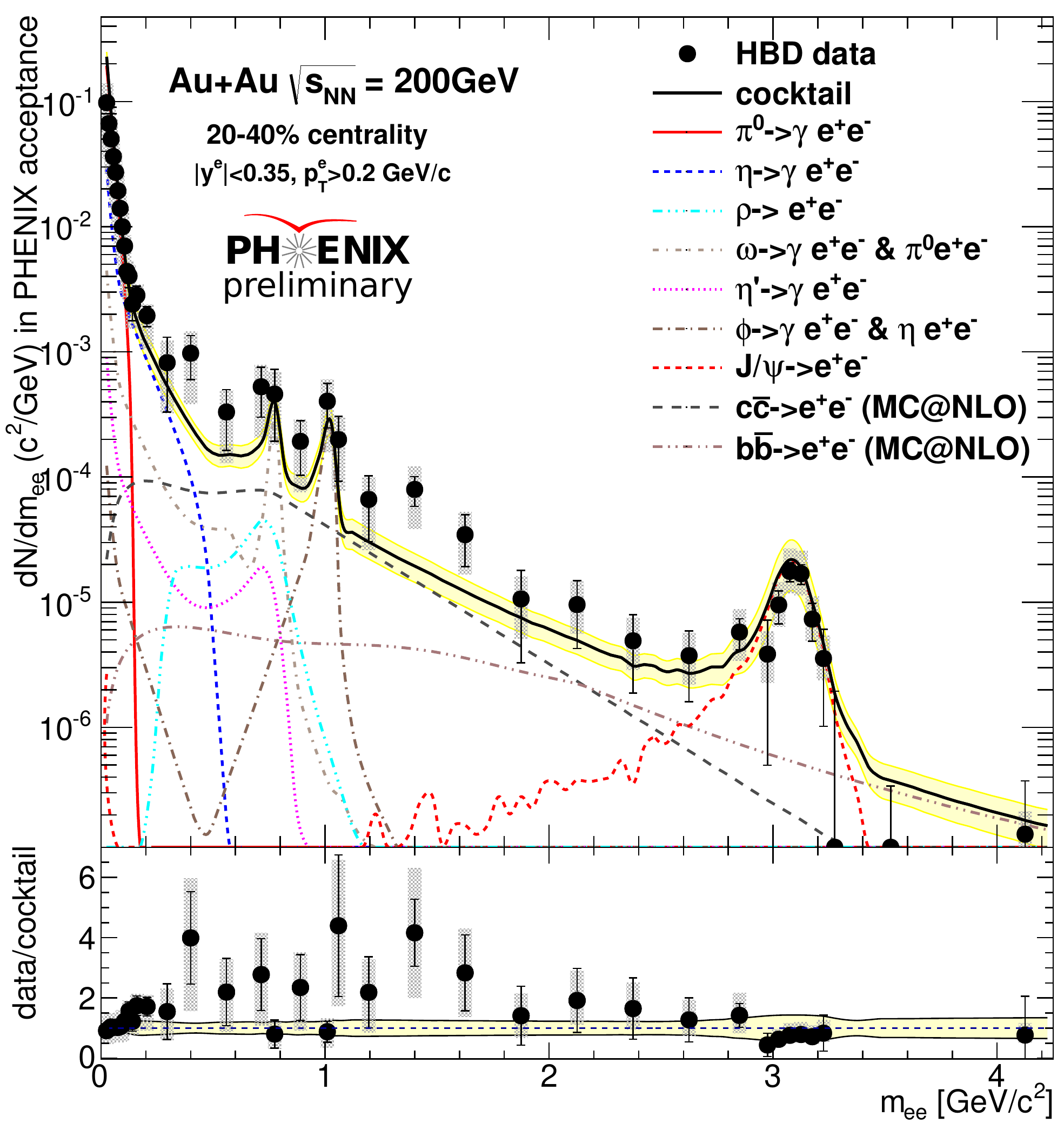}
\end{minipage}
\begin{minipage}{0.33\linewidth}
\includegraphics[width=0.9\textwidth]{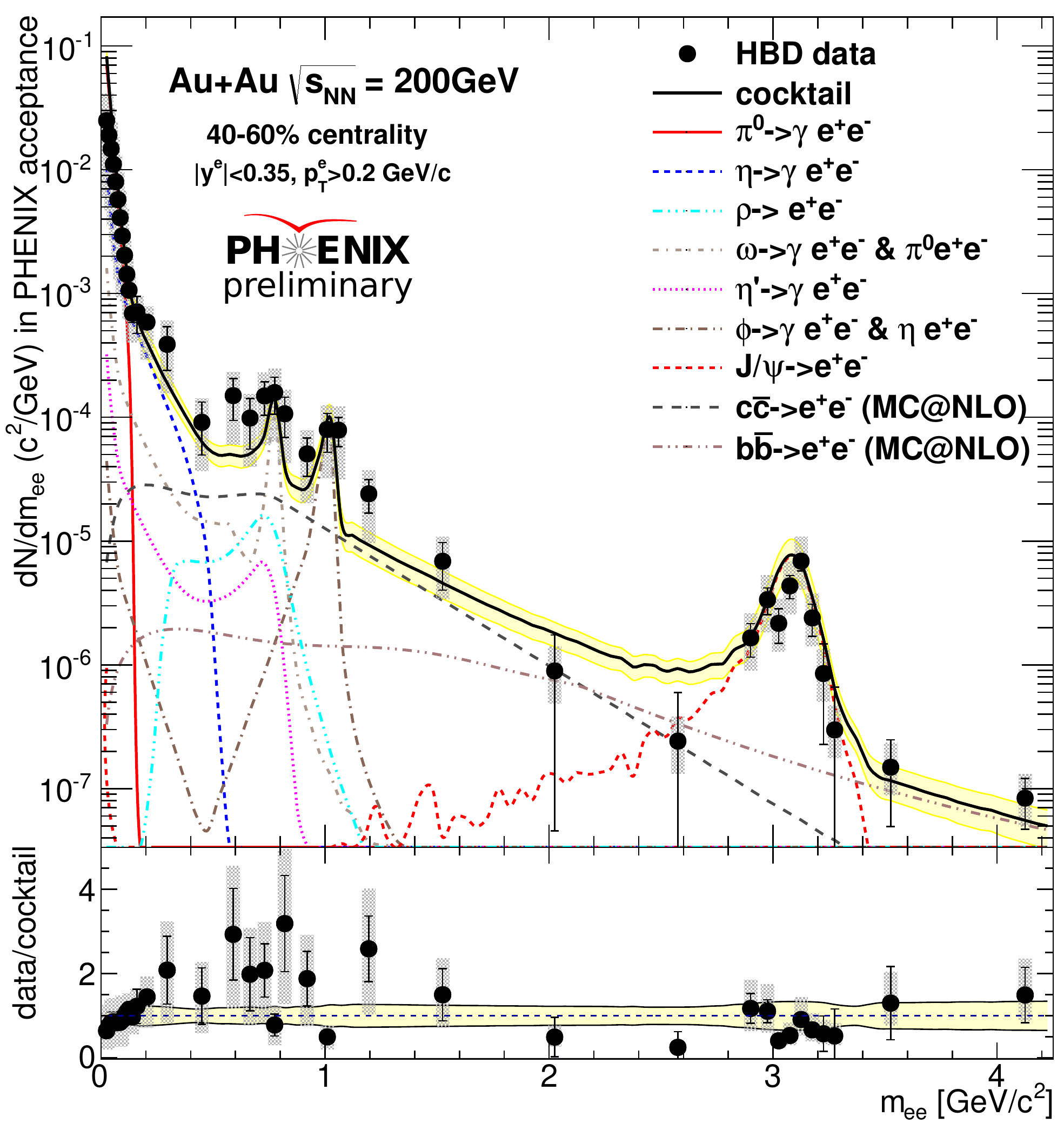}
\end{minipage} 
\begin{minipage}{0.33\linewidth}
\includegraphics[width=0.9\textwidth]{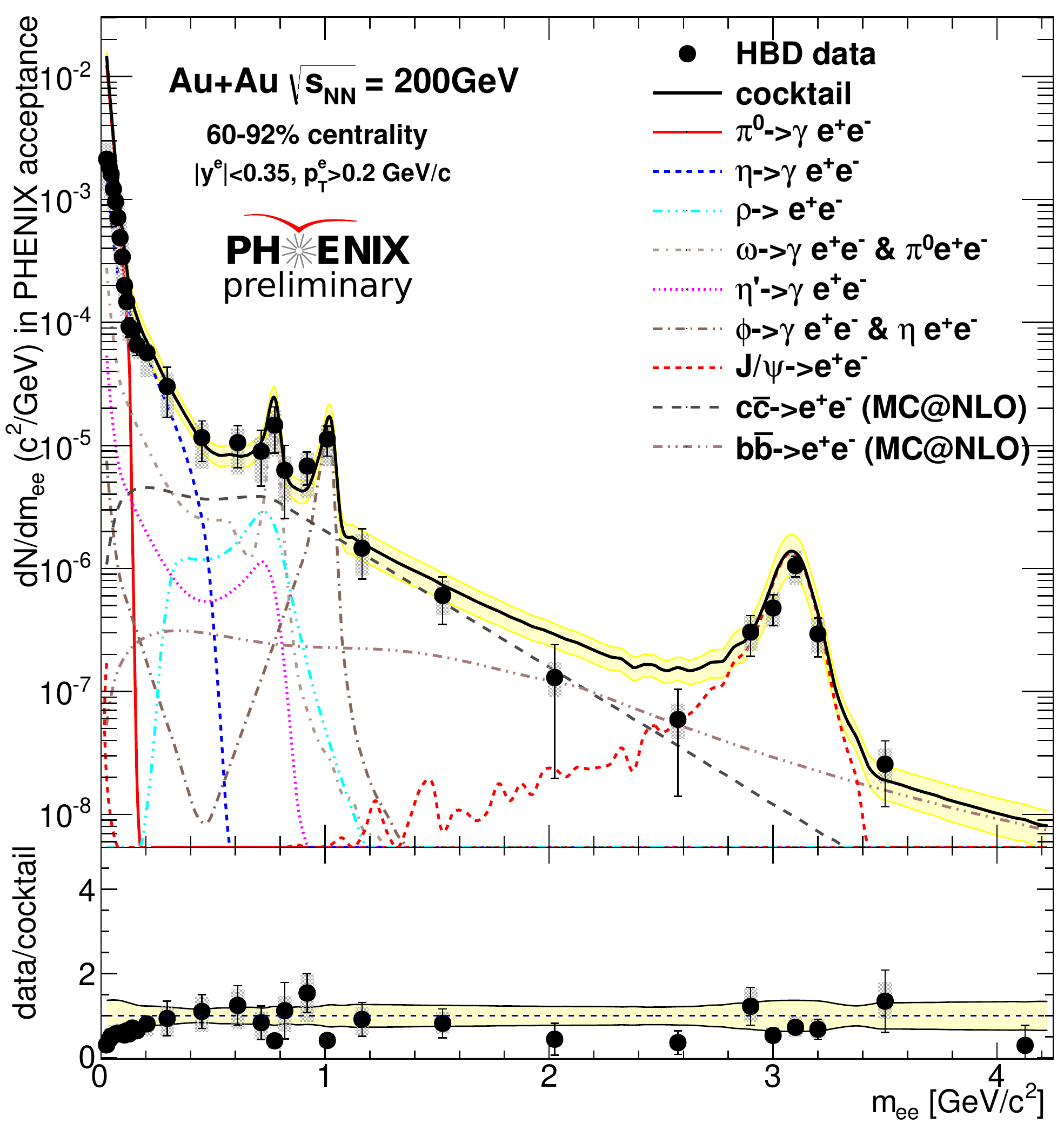}
\end{minipage} 
\caption{\label{fig:compare_auau}Dielectron mass spectra from  Au+Au collisions for three centrality bins 20-40\% left, 40-60\% center, 60-92\% right).  The ratios to the cocktail are also shown in the lower panels.}
\end{figure}

Two neural networks were utilized to improve the electron ID and single/double electron separation. The neural networks were trained using detailed GEANT simulations. For the first stage of the analysis, a neural net selects tracks with a high probability for being an electron where studies have shown that the purity of the electron sample approaches $90\%$. A second neural net is used to reject contributions from double electrons in the HBD. This has been shown to remove up to 70\% of double electrons reconstructed as singles by the central arm.\cite{Makek}

The background subtraction technique used for Au+Au is slightly more complicated than that used for $p$+$p$ due to the larger combinatorial background in central events. There are two steps to remove the two types of background. First, a mixed event subtraction is used to remove the combinatorial background and then a like-sign subtraction is used to account for the correlated background. To isolate the correlated spectra from the like-sign pairs, background is subtracted using mixed events and the remaining correlated like-sign pairs are then corrected for relative acceptance, where FG12 is the unlike-sign foreground, BG12 is the unlike sign mixed event background, and FG(LS and BG(LS) are the foreground and background for the like-sign respectfully

\[S=FG12 - N\times BG12 =\alpha \times (FG(LS) - N \times BG(LS)) \]

 The normalization is done after a pair cut efficiency is applied and a normalization factor is calculated such that the relationship $\langle BG12\rangle=2\sqrt(\langle BG11\rangle \langle BG22\rangle)$ holds. The results are shown in Fig. 3 for 3 different centrality bins. In Fig. 4 these results are compared to the Au+Au obtained in Run 4 before the HBD was installed.\cite{Tserruya} 




\begin{figure}[ht]
\begin{minipage}{0.5\linewidth}
\includegraphics[width=0.9\textwidth]{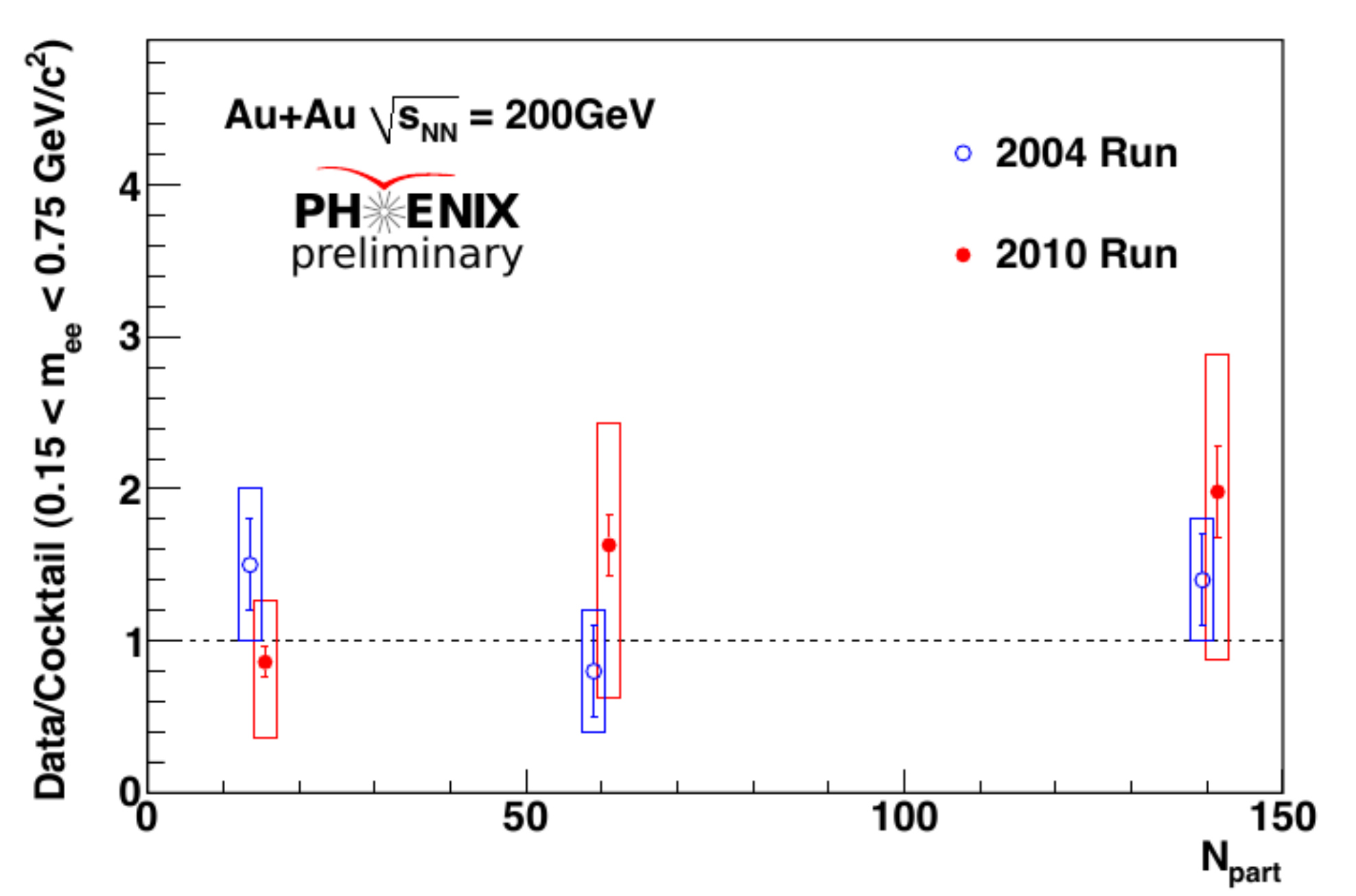}
\end{minipage}
\begin{minipage}{0.5\linewidth}
\includegraphics[width=0.9\textwidth]{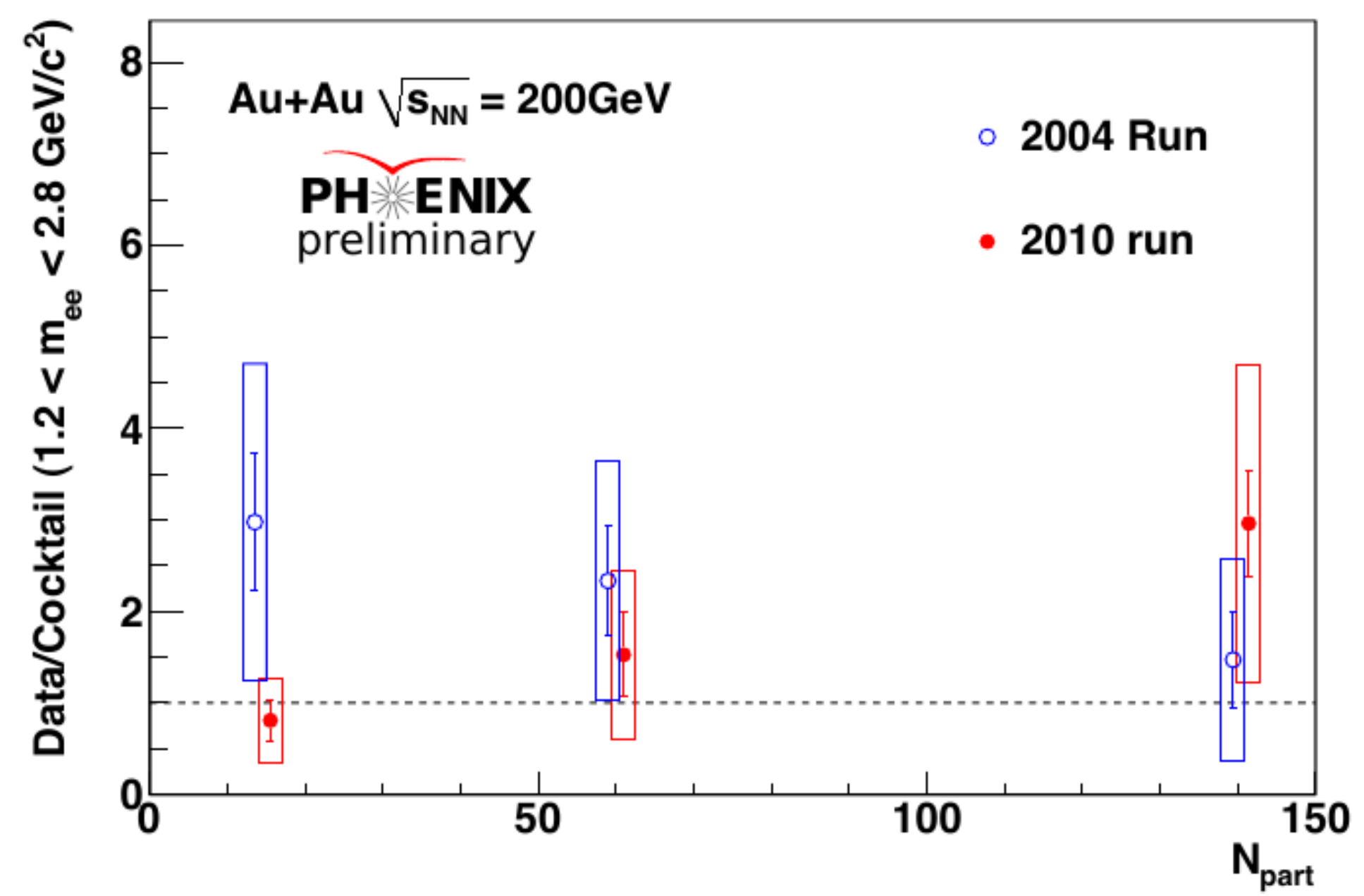}
\end{minipage} 
\caption{\label{fig:compare_auau}Comparison of Au+Au results to previous published results. We see that the latest results using the HBD are fully consistent with previous results within the systematic errors. The ratios are plotted as a function of Npart integrated over two different mass regions.}
\end{figure}

\section{Summary}

The dielectron mass spectrum has been presented for $p$+$p$ and for three centrality bins of Au+Au. These results are consistent with previous published results at PHENIX. There is a hint of enhancement in the 20-40\% centrality bin in the low- and intermediate-mass region. Currently this is not yet conclusive given present uncertainties and work is in progress to improve the systematics.

\end{document}